# Vishap epoch unitary society in Armenian Highlands, c. 4000 BC: data analysis consequences

Vahe Gurzadyan[1,*] & Arsen Bobokhyan[2]

[1] Center for Cosmology and Astrophysics, Alikhanian National Laboratory and Yerevan State University, Yerevan, Armenia

[2] Institute of Archaeology and Ethnography, National Academy of Sciences and Yerevan State University, Yerevan, Armenia

* Author for correspondence gurzadyan@yerphi.am

**Abstract.** Vishaps—dragon stones—discovered in the Armenian Highlands convey a remarkable message about the spiritual and social character of their epoch, c. 4000 BC. The unexpected bimodal distribution of their elevations indicates the deliberate, labor-intensive placement of these massive stones—some weighing up to 7–9 tons—in locations where the period suitable for construction activities at high altitudes was extremely limited. Their positions, correlated with nodes of previously identified prehistoric irrigation systems, support the interpretation that they were dedicated to a cult of water. This evidence points to the existence of an organized and unified society capable of sustaining and maintaining such a resource-intensive cult.

## Introduction

In a recent study[1], the first statistical analysis of vishaps—"dragon stones," unique prehistoric archaeological monuments of the Armenian Highlands (Armenian *vishap* meaning "dragon")—was conducted. The bimodal distribution of their elevations was unexpectedly found to indicate that their construction was intentionally labor-intensive rather than arbitrary, supporting the hypothesis that they were dedicated to a water cult.

In the present study, we continue the analysis of vishap data, further elucidating their cultic significance. We conclude that these monuments convey a message about the existence of a unitary society in the Armenian Highlands during their period. Their placement at critical points of ancient irrigation systems links the cultic function of the monuments with the constructional and organizational capabilities of that society[2,3].

Vishaps are stelae ranging from two to five meters in height, made of volcanic rocks (andesites and basalts). They are unique among Ancient World stelae due to their exceptional animal iconography and distinctive placement in the landscape (Figs. 1, 2). Research on vishaps has been ongoing since the early 20th century, increasingly incorporating scientific methods and detailed landscape analyses.

Morphologically, vishaps appear in three main types: the *piscis* type, sculpted in the form of a fish; the *vellus* type, carved to resemble a bovine hide draped with head and extremities; and the *hybrida* type, which combines elements of both iconographies. Typically, they are situated in secluded, water-rich, high-altitude meadows within unforested mountainous regions, at elevations of approximately 1,200–3,200 meters above sea level (asl).

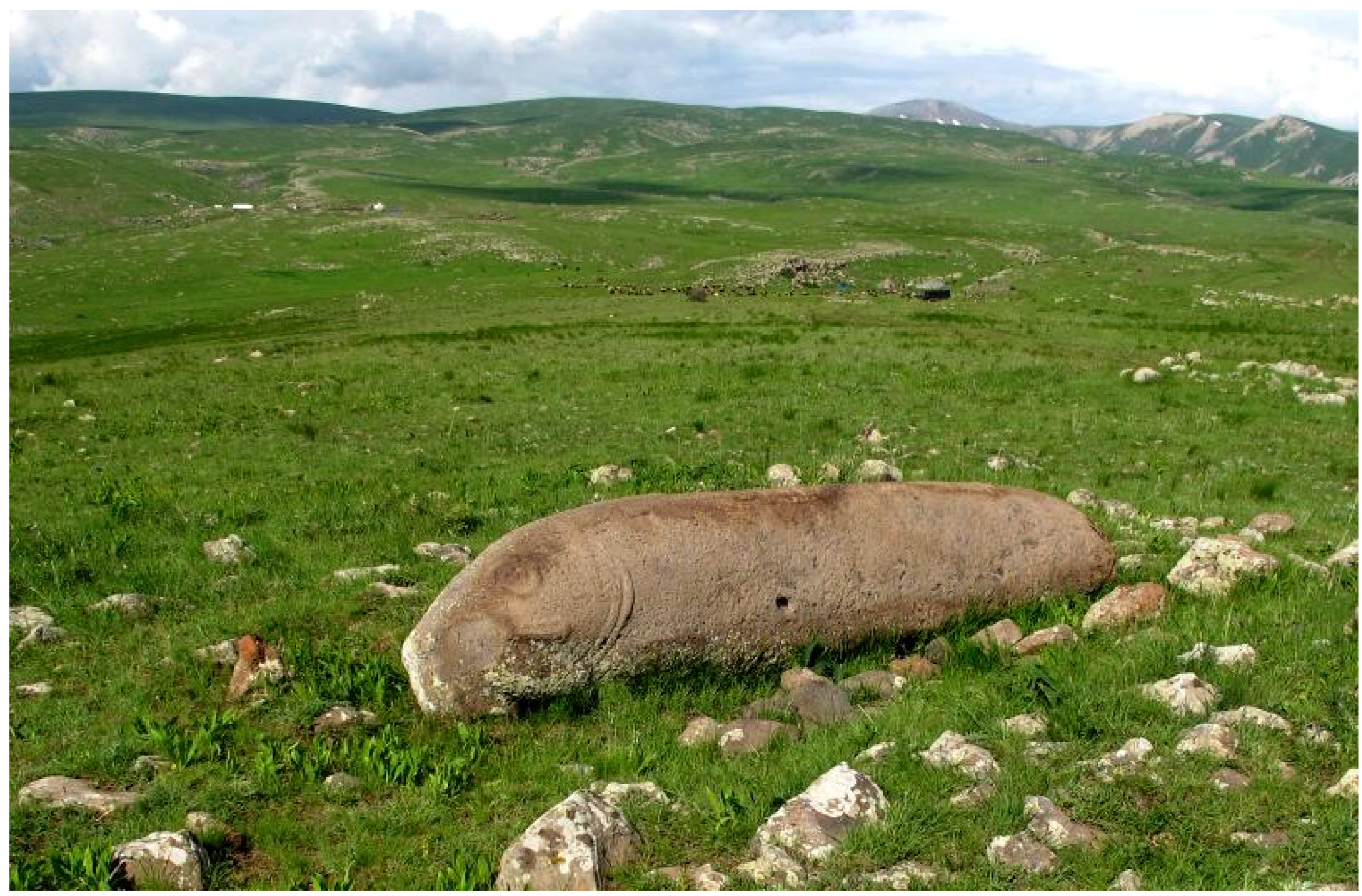

**A.**

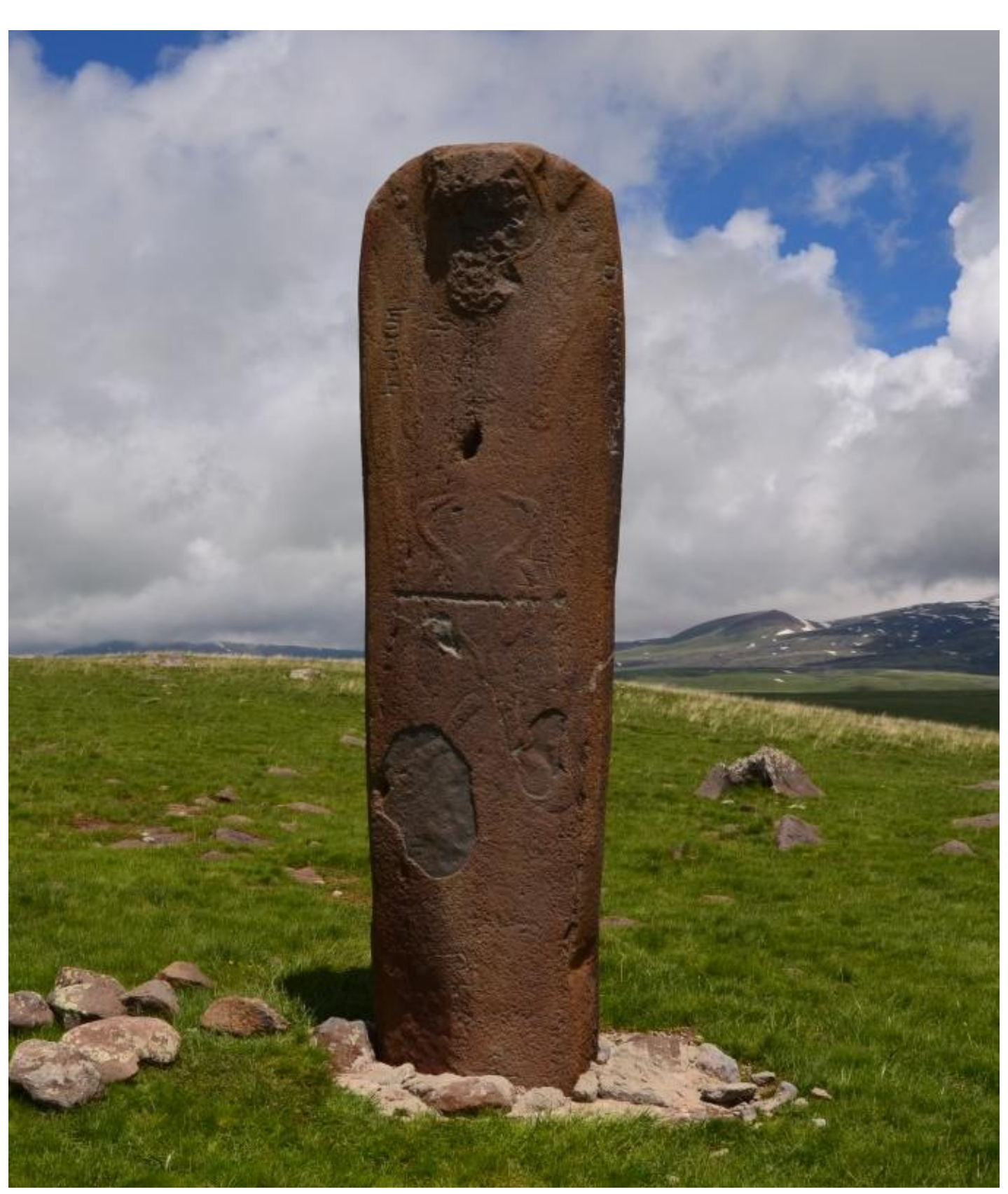

**B.**

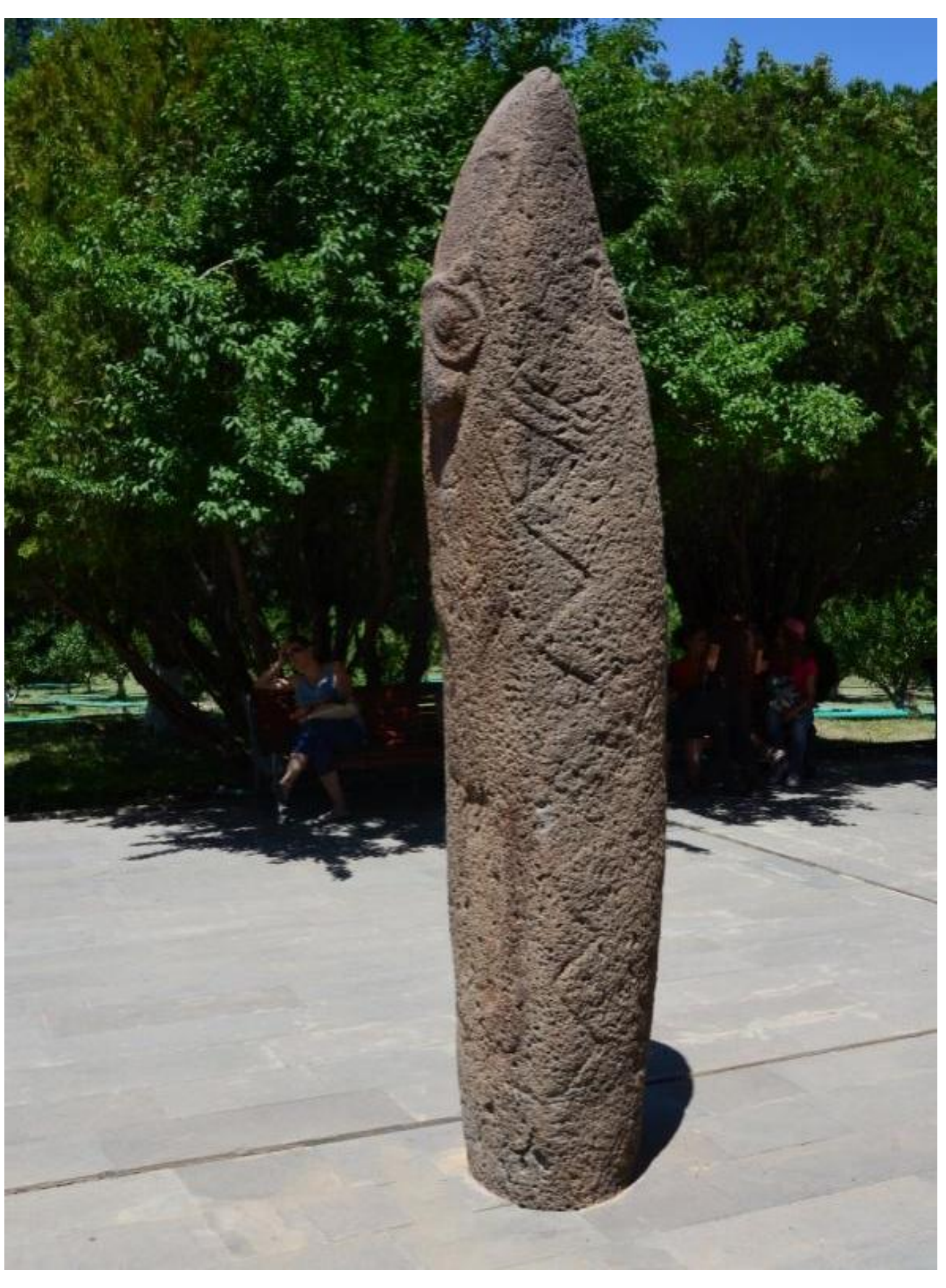

C.

Fig.1. **Examples of vishap stelae**: **A**. piscis (Vanstan 2), **B**.vellus (Gegharda lich 1), **C**. hybrida (Sakhurak 5) (photos by A. Bobokhyan).

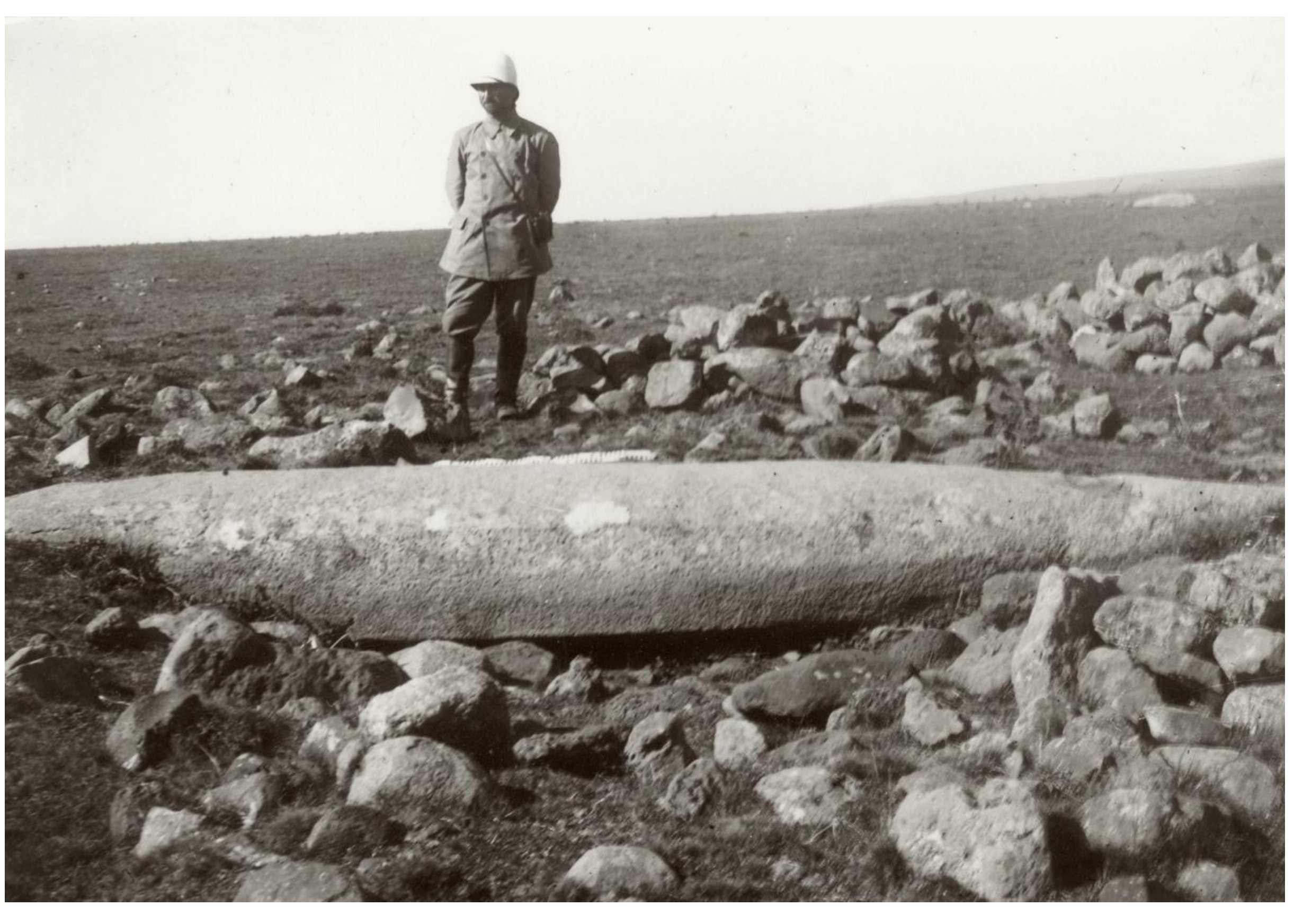

Fig. 2. Archaeologist Ashkharbek Kalantar in 1927 near the vishap Karakap 1, Mt. Aragats (photo from ref.[9]).

The macro-regional distribution of vishaps extends from the Lake Van region in the south, to the Trialeti mountain range in the north, the Erzurum region in the west, and the Sevan mountain range in the east. Their densest concentrations are found on Mt. Aragats and in the Geghama mountains of Armenia. To date, approximately 115 vishaps have been documented within the territory of the Republic of Armenia.

The *vellus* vishaps are the most numerous and are distributed fairly evenly across the macro-region. *Hybrida*-type vishaps are the rarest and, so far, have been found only in the southwest of the Geghama Mountains in Armenia. *Piscis*-type vishaps are currently absent from the westernmost parts of the macro-region. Because the iconography of vishaps is unique, standardized, and highly specific, their broad distribution indicates the presence of a shared symbolic and religious framework across the region at the time of their erection[1].

The most significant high-altitude site featuring vishaps and related archaeological monuments is Tirinkatar, a site covering over 40 hectares on the southern slopes of Mt. Aragats, at an altitude of approximately 2700–3100 m. Locally, pastoralist communities know it as “Karmir Sar” (Armenian for “Red Hill”). The site is a stunning summer pasture, rich in grass and water, with spectacular views of both the Mt. Aragats summits and Mt. Ararat. Tirinkatar was discovered in 2012, and systematic surveys and excavations have been ongoing since then[4-6].

Radiocarbon dating indicates that the erection of two vishaps at the site corresponds to roughly 4200–4000 BC. Archaeological evidence also shows that Tirinkatar was used as a campsite at least from the Neolithic period, i.e., from the end of the 6th millennium BC onwards. Between the 3rd and the end of the 2nd millennia BC, additional features were added, including large aggregated cell structures, circular stone structures commonly termed “cromlechs”, circular structures with stone fillings and inner chambers known as “giant’s houses”, as well as barrows and petroglyphs.

Vishaps represent the earliest known examples of figurative monumental art in the Armenian Highlands and the Caucasus. Elaborated on four sides, they can also be considered among the earliest monumental statues in human history. Their uniqueness is evident both in their geographic and altitudinal placement and in their iconography, which is highly specific and has no known parallels in the ancient world. Their association with the cult of water highlights their complex spiritual significance.

The Tirinkatar[4–6] site contains the largest concentration of vishaps—12 in total—and is possibly one of the highest-altitude archaeological complexes in the world. It is an exceptional example of early seasonal settlement and the domestication of high mountainous areas, reflecting an organized and continuously evolving cultural landscape. Comparable high-altitude sites in other regions, such as the Andes and the Himalayas—like Machu Picchu, Tiwanaku, and Chavín de Huantar—are primarily dated from the Late Bronze Age to the Medieval period. By contrast, Tirinkatar was in use as early as the Neolithic period.

The archaeological site clusters around the vishaps, together with other features such as ritual platforms, giant's houses, cromlechs, and petroglyphs, represent the earliest and most distinctive cult systems typical of the Armenian Highlands. The Tirinkatar complex vividly demonstrates how spiritual practices—and, consequently, the management of collective memory—influenced the formation of early identities and became a primary means of uniting communities and organizing economic activities, including irrigation, herding, and seasonal migrations. As the main cult center, Tirinkatar, along with its vishaps, links the earliest monumental art, a cultic site, and evidence of seasonal pastoralism in the high mountains—not only within the Caucasus but possibly beyond.

The exceptionally high density of archaeological features still visible on the surface reveals continuity of human activity over millennia, reflecting the existence of a rich cultural landscape in these highlands. In the following analysis, we examine the vishap data in the context of cultic phenomena, particularly their relationship to ancient irrigation networks, and highlight the remarkable message these prehistoric monuments convey: the presence of a cohesive society capable of sustaining and serving such a resource-intensive cult.

## Methods

### Cult as resource-consuming capacity of unitary societies and labor-intensive vishaps as indication of a cult

The concept of labor was used[1] as an anchor in the first statistical analysis of the 115 vishap data in order to investigate the motivations behind their construction. Specifically, the sizes of the vishaps and the elevations of their mountain locations were analyzed as factors linked to the amount of labor required:

(a) The larger the vishaps, the greater the labor investment needed for their production, including the extraction of the stone from rock formations, polishing, and transportation to the intended locations;
(b) The higher the elevation, the shorter the available snow-free period for performing the labor necessary to produce and transport such heavy objects.

The unexpected bimodal distribution of altitude identified[1] indicates an intentional labor-intensive motivation for their construction and placement at high altitudes in areas of primary snow storage. This observation supports the interpretation that they were dedicated to the cult of water, particularly in view of their specific spatial patterns, discussed below, which are associated with water sources, nodes of ancient irrigation systems, and artificial reservoirs.[7-12]

The analysis presented below of vishap altitude and size distributions is based on the sample of 115 items listed in[1]. The objective of the present study, within the framework of the statistical methodology applied to the vishap data[1], is to examine the stability of the bimodal distribution—that is, to assess its sensitivity to the two key parameters of the vishaps: their size and their elevation. These parameters are directly related to the concept of labor consumption. Accordingly, we analyze several subsamples of the dataset, defined by specific constraints on size and altitude, in order to observe how the bimodal distribution behaves under these conditions.

## Results

### Vishap altitude bimodal and size distributions

In Figure 3, the altitude distribution (in meters) is shown for all 115 vishaps and for the 65 specimens larger than 300 cm. Both distributions display a bimodal pattern, indicating that the larger, more labor-intensive vishaps were predominantly situated at higher altitudes. This placement required extra effort for both their creation and transport, given the shorter snow-free season at higher elevations. The additional effort encompassed person-day labor under harsh high-altitude conditions as well as logistical organization to supply workers with food, fuel, and transport.

Figure 4 presents the size distribution (in centimeters) for all 115 vishaps and separately for those located above 2000 m and 2500 m asl. The data show that vishap sizes do not decrease with increasing altitude, despite the higher labor and organizational demands.

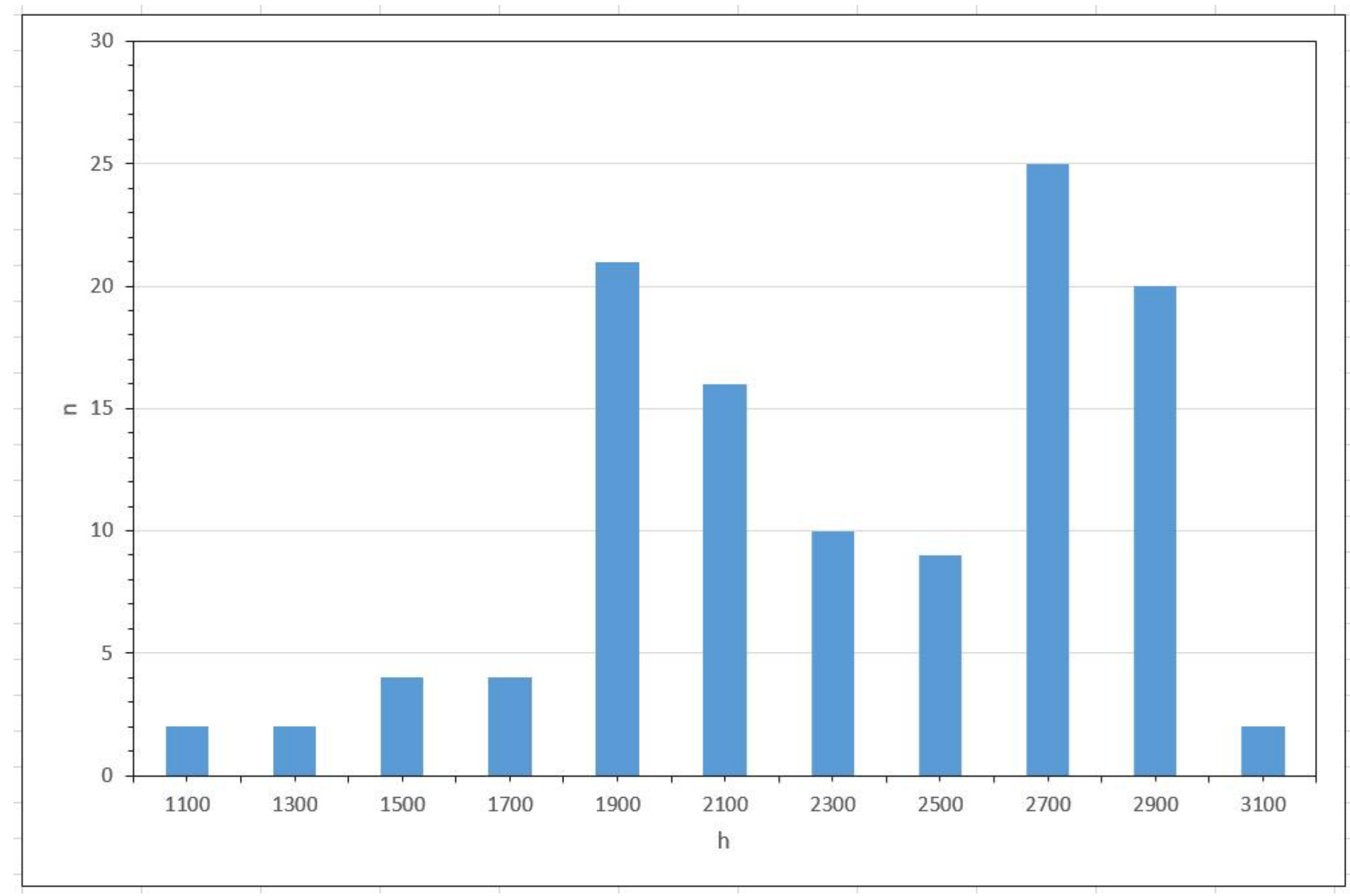

**A.**

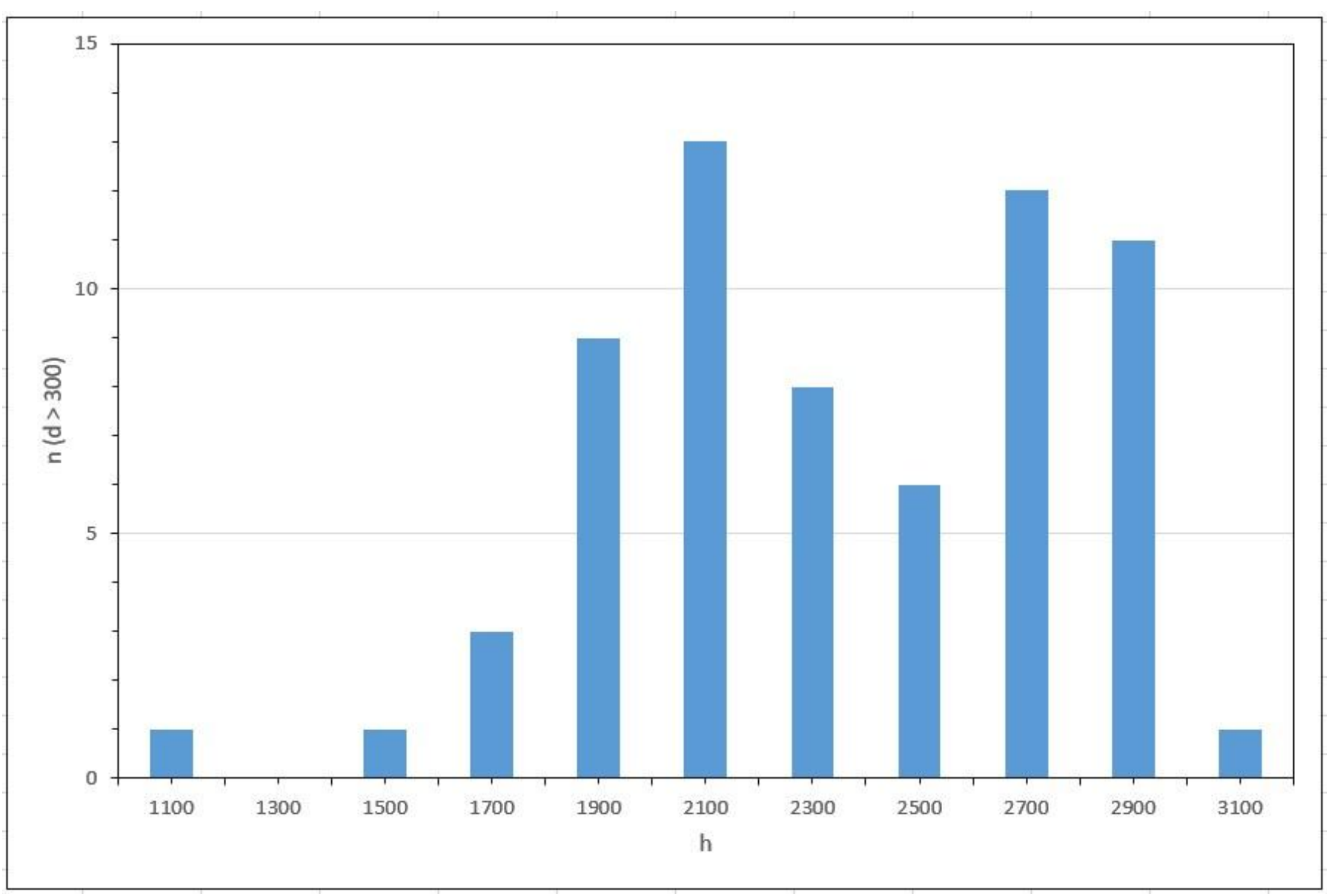

**B.**

Fig. 3. **Vishap altitude distributions. A.** Distribution of the location altitudes for 115 vishaps of all sizes; **B.** the same for those of sizes bigger than 300cm (by V. Gurzadyan).

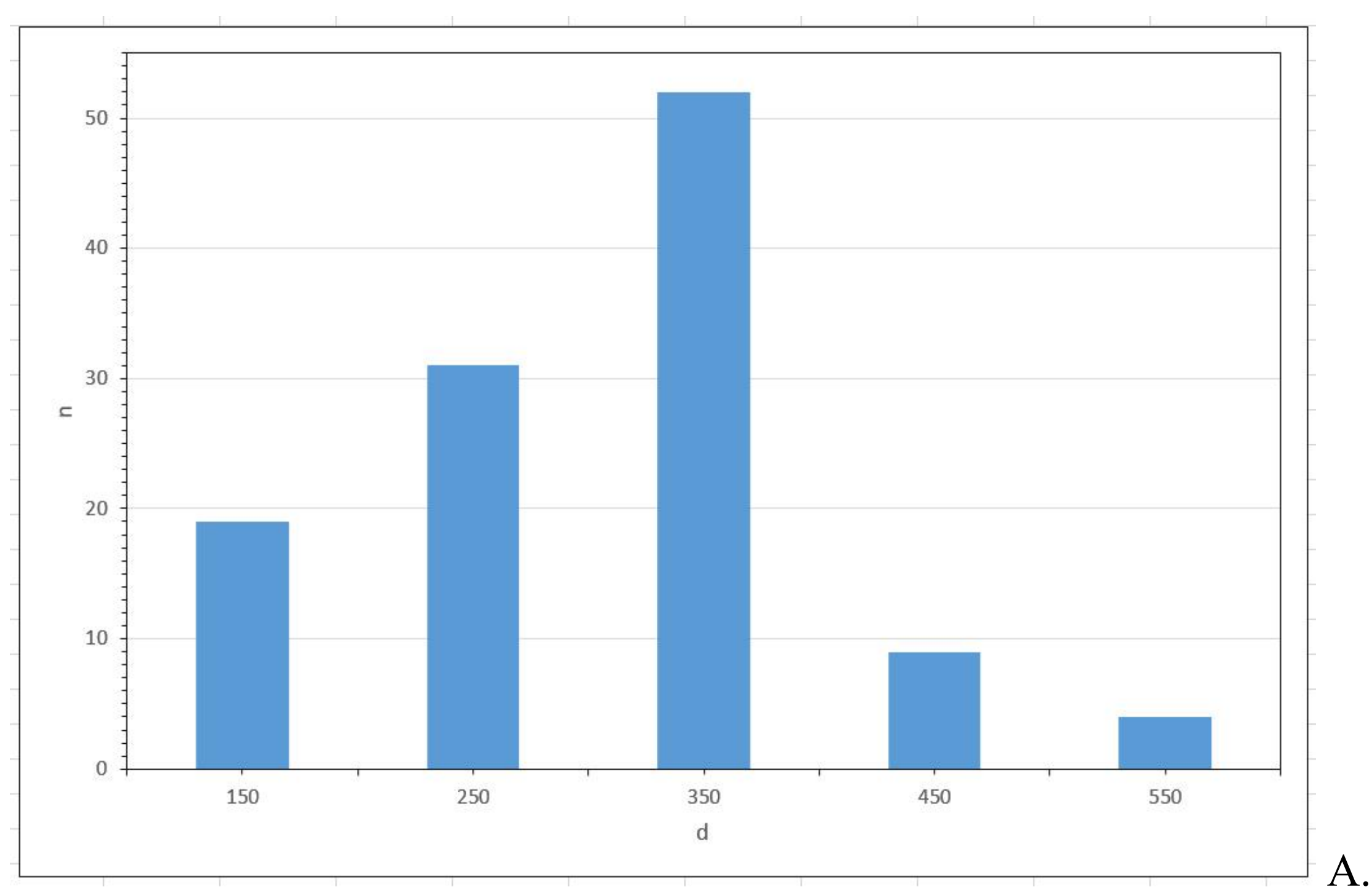

0
10
20
30
40
50
n
150
250
350
450
550
d

A.

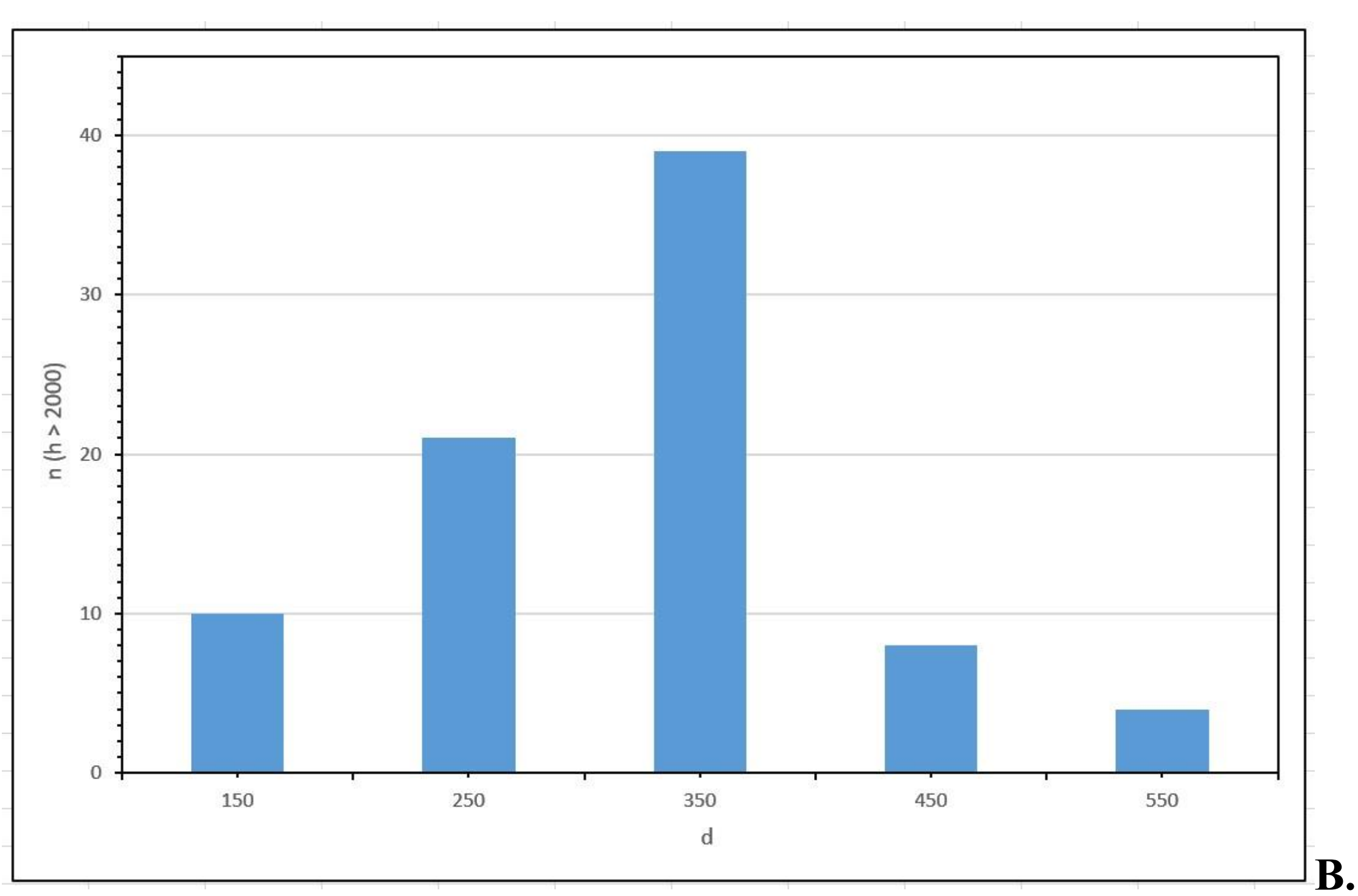

0
10
20
30
40
n (h > 2000)
150
250
350
450
550
d

B.

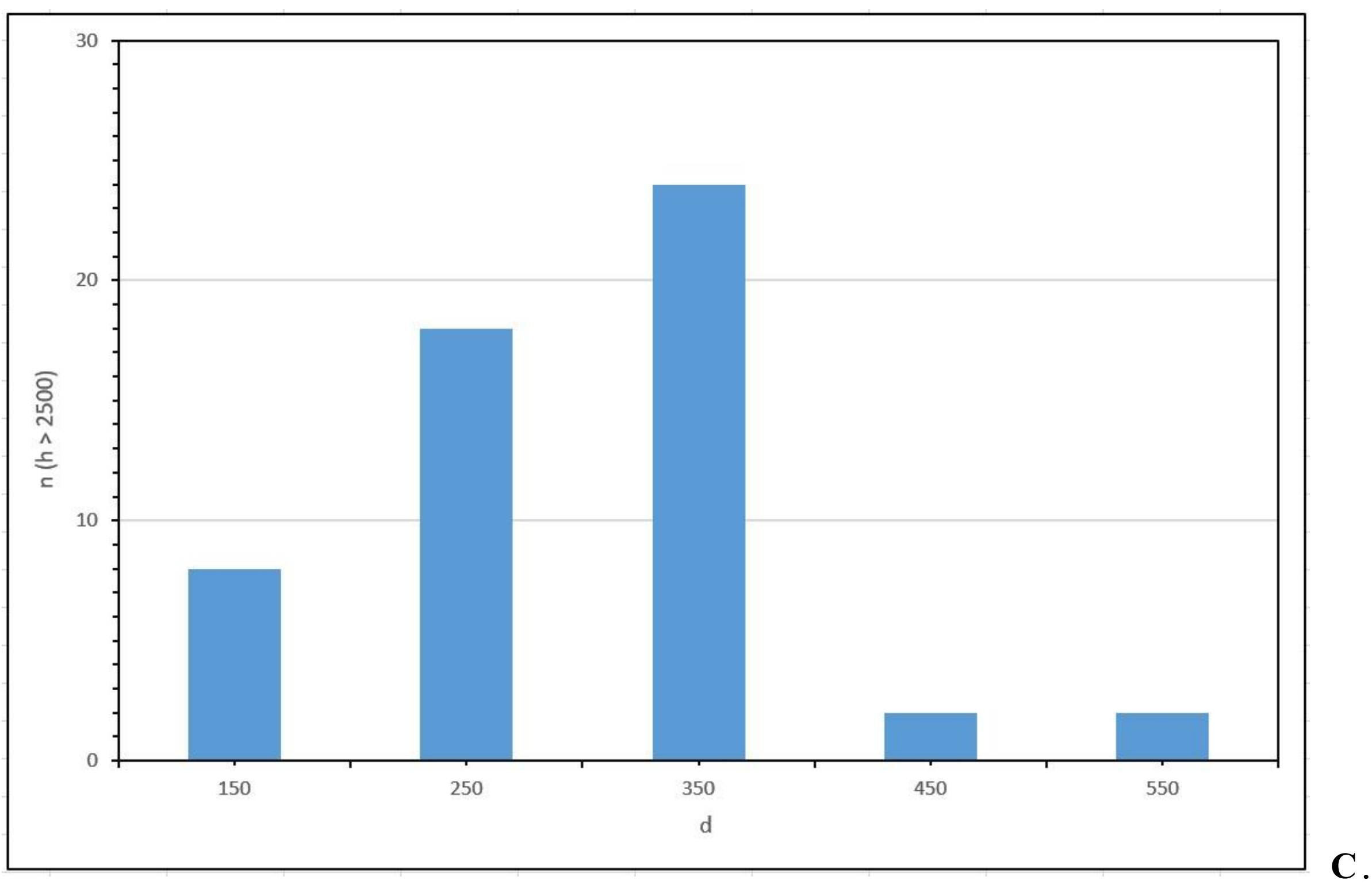

Fig. 4. **Vishap size distributions. A**. The distribution of 115 vishap sizes; **B.** The same for those (82) located at altitudes higher 2000m, asl; **C**. The same for those (54) located higher 2500m, asl. (by V. Gurzadyan).

The persistence of the bimodal pattern in the larger subset of vishaps, as well as among those located at higher altitudes, strongly reinforces the notion that their construction was driven by cult motivations, involving organized, labor-intensive efforts, rather than purely practical motives.

## High-Altitude Irrigation Systems as Indicators of Unitary Societal Efforts

In the context of the current discussion, it is essential to examine the prehistoric water distribution systems of the Armenian mountains. Detailed fieldwork documentation by Ashkharbek Kalantar[7-12] in the Aragats and Geghama Mountains during the 1920s–30s led to the development of a perspective suggesting that the *vishap* stelae, dating to the prehistoric period, were strategically located at key points of the irrigation systems supplying water to the Ararat Plain (Figs. 5–7). This placement indicates a direct connection between these stelae and the water networks. Consequently, the *vishaps* have been interpreted as symbols of water worship, guardians of water, and patrons of canal construction. This association with the social developments of ancient Armenia[13,14] has become a dominant view among subsequent researchers[15,16].

Unlike the *vishap* stelae, which can be clearly dated to the prehistoric period, the chronology of the water distribution systems remains ambiguous, particularly due to their continuous use and renovations across later epochs[17-20]. Nevertheless, analysis of the archaeological context allows for a tentative correlation between the stone stelae and the irrigation networks, reflecting a certain degree of synchronicity between them.

First, this concerns the patterns of placement of the discussed units. The ancient irrigation systems of Mount Aragats (Figures 5, 6) and the Geghama Mountains are characterized by a complex network of artificial ponds and canals extending across the slopes and foothills of these mountain ranges. Their primary purpose was to support agriculture and horticulture in the Ararat Plain, as well as animal husbandry, by providing drinking water. Snowmelt and rainwater flowing from the mountains in spring were harnessed for this purpose.

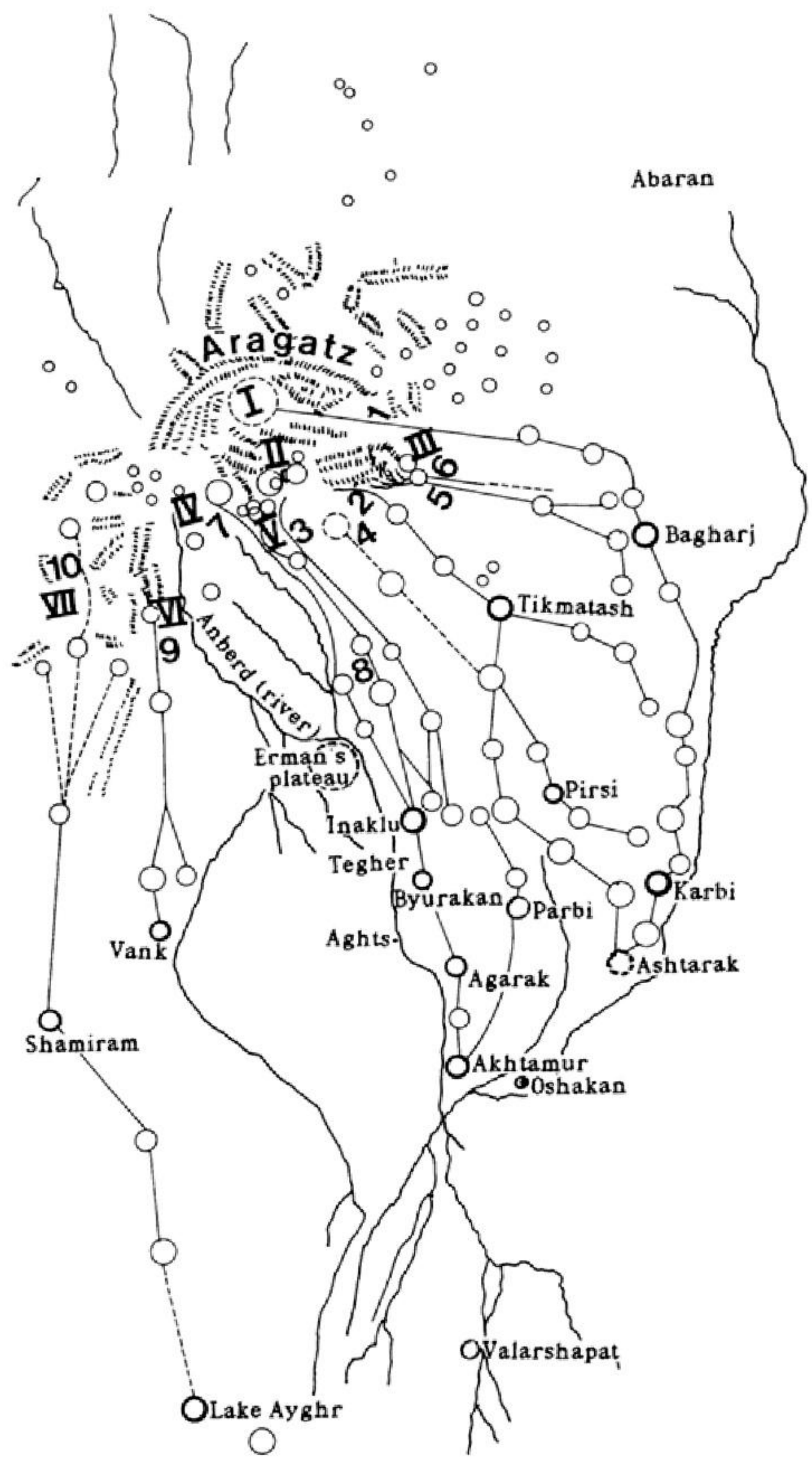

Fig. 5. The ancient water distribution system map of Mt.Aragats by Ashkharbek Kalantar (from ref.[10,11] ). Shamiram – modern toponyms; II – Latin numerals denote water collector reservoirs; 3 – Arabic numerals denote main canals; o – round points denote ponds.

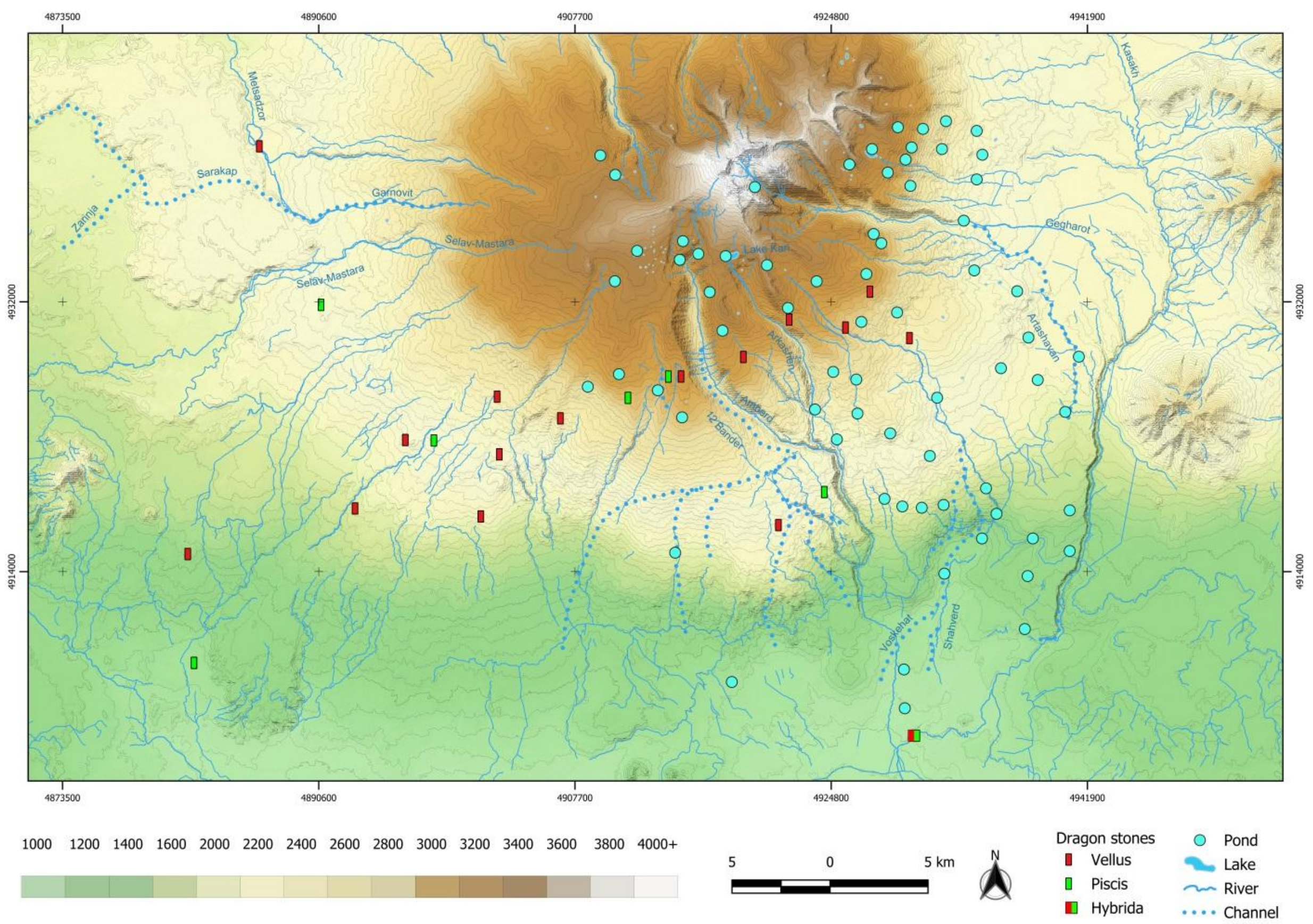

Fig. 6. Modern reproduction of Kalantar's 1937 map (Fig.5) of the ancient water distribution system of Mt.Aragats with location of currently visible landscape units and vishap stelae (map by A. Bobokhyan, drawn by A. Ananyan).

Within these ancient irrigation networks, two types of reservoirs or ponds can be distinguished based on their location and function:

*Water collection ponds*, situated in the upper parts of the network, which store meltwater and feed the canals originating from them. *Pond-reservoirs*, aligned along the canals and characterized by artificial embankments. These feed *primary canals,* which originate from the water collection ponds, and *secondary canals*, which branch from lower ponds or from the main canals to irrigate specific areas. Notably, the vishap stelae are typically located at the converging points of these irrigation systems[9,10].

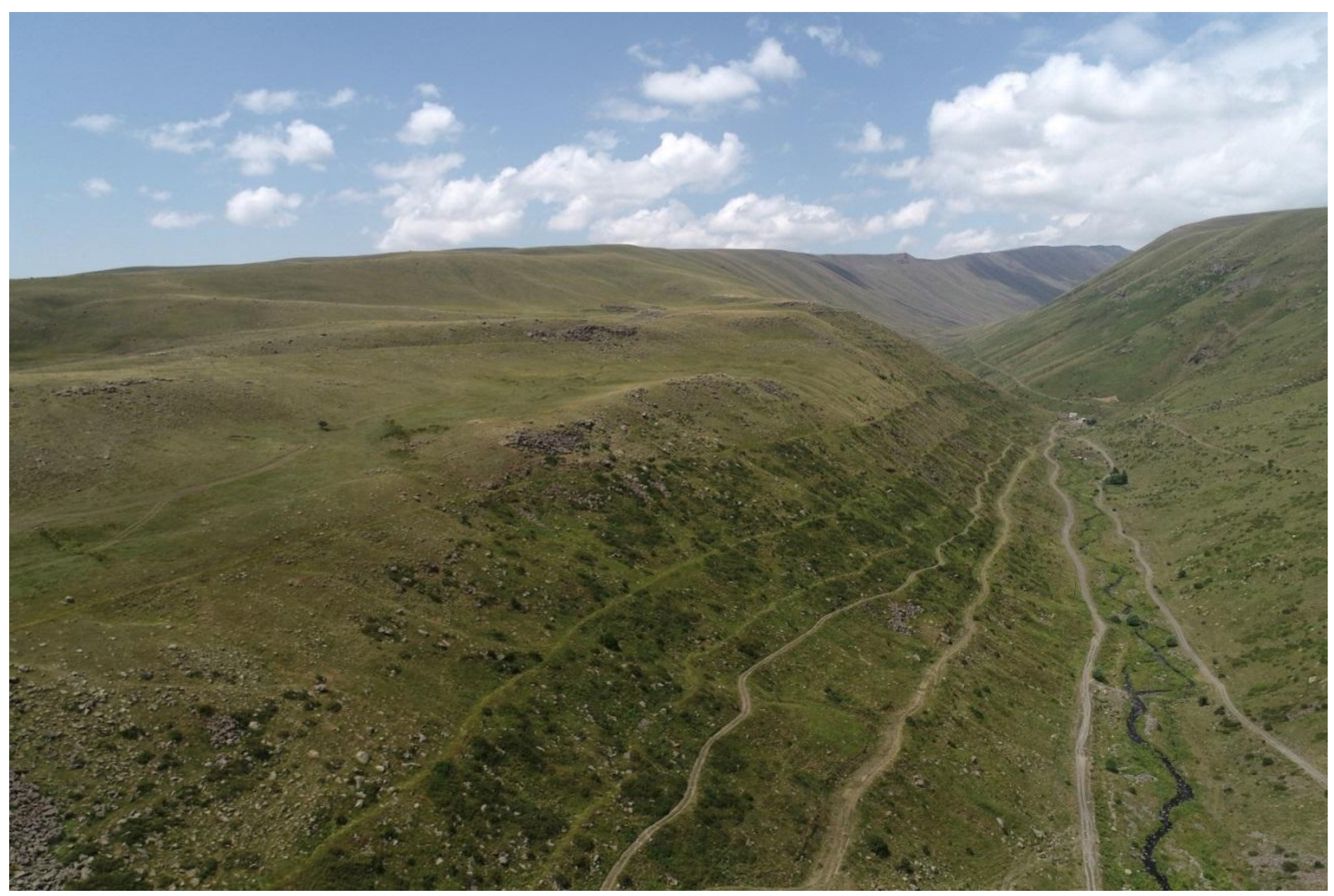

Fig. 7. The 12 Bander canal system south-east of Tirinkatar (photo by A. Bobokhyan).

Second, consider the stability of the water distribution systems. Across nearly all periods, these systems have remained remarkably uniform, operating according to consistent mechanisms and principles (ref. [9], p.54). Consequently, several contemporary water structures on Aragats trace their origins back to the prehistoric period, albeit with subsequent reconstructions and engineering modifications. For instance, a section of a cyclopean masonry dam near the outlet of Lake Kari on Aragats (3200 m asl) preserves evidence of early hydraulic engineering[10]. A similar pattern of stability is observed in the Geghama Mountains, particularly regarding Lake Geghard, whose reservoir was restored in the second half of the 19th century by the Geghard Monastery.

This enduring stability suggests a continuous development of water distribution systems from the Early through Late Bronze Ages[21-23]. Within this continuity, the Urartians expanded and modernized existing water networks using advanced engineering techniques, often marking the headwaters of canals with inscriptions and dedication stelae[24]. While such stability complicates the precise dating of canals, it simultaneously allows for the study of these systems in *longue durée*, highlighting recurring features and similarities characteristic of different periods.

Large-scale enterprises, such as the construction of huge irrigation networks, along with the relevant cults, are certainly shaping the organizational level of the societies. Water networks then link the utilitarian and spiritual -- vishaps as their keepers -- aspects of the society activities. Then, vishaps, with their certainly not accidental concentration of Mount Aragats slopes facing the Ararat valley, represent the significant information carriers on the archaic epochs. In this context, the approaches to the more precise dating of the monuments and of the sites[25-29], including laboratory methods, have to be continued.

## Discussion

It is noted[1] that, as human history reveals, usually the cults are indeed associated to significant resources (labor) provided by their societies. Those resources typically involve non-utilitarian goals -- huge monuments, pyramids, temples -- and by the rulers -- pharaohs, emperors -- are justified to their societies within a cult as absolutely necessary price to their utilitarian problems.

In the conditions of the Armenian Highlands, characterized by relatively dry and hot summers in the valleys, the vast amounts of snow accumulated in the mountains serve as the primary source of water for springs and rivers throughout the year, forming a fundamental basis for life and prosperity.

The remarkable labor and organizational effort required for construction using prehistoric technologies, as well as for transporting the heavy vishaps to high-altitude areas associated with these life-giving water sources, appears to have been a well-justified and communicated endeavor by the rulers to their societies. For instance, the Karakap 1 vishap on Mt. Aragats (Fig. 2), carved from basalt and situated at an altitude of approximately 2800 m, measures 506 × 95 × 65 cm and weighs around 7 tons (for the definition of volcanic rock vishaps and their source identification, see[30]).

The bimodal distribution of vishap altitudes indicates a deliberate choice of high-altitude placement for these labor-intensive monuments. This is further corroborated by the distribution of their sizes, which shows no decline in the number of large items located at higher elevations.

Thus, a deliberate resource-consuming and non-utilitarian activity revealed by the vishap parameters indicate that we deal with a phenomenon of a water cult. Moreover, the correlation of the vishap location patterns with the developed irrigation networks on Mount Aragats and Geghama Mountains is linking the spiritual and utilitarian activities, outlining the organizational skills of the ancient societies in c.4000 BC.

Indeed, large-scale irrigation systems are typically associated to unitary societies, and in many cases it is precisely through the construction and management of such systems that such organized social structures emerge[31-34.] This pattern is particularly evident in the Ancient Near East, especially in Mesopotamia, where the earliest irrigation installations appear as early as the fifth millennium BC, e.g., at Choga Mami[35,36] and continue to be documented through the mid–first millennium BC[37-40]. Although the organization of irrigated landscapes varied between terraced mountain zones and the open steppe lowlands, the outcomes were comparable—both settings reveal the emergence of increasing social organizational level.

We can conclude that only a unified and sufficiently organized society would have been capable of supporting the resource-intensive cult described above and managing extensive mountain-coverage irrigation works. This is the precious message that silent monumental sculptures have transmitted to us over millennia.

The *vishaps* of the Armenian Highlands serve as exemplary carriers of the spiritual and socio-organizational legacy of prehistoric societies, particularly in the context of specific environmental conditions[41,42], such as high-altitude archaeological sites in the Andes[43-46] or the moai of Easter Island[47.]

**Acknowledgements**

For essential contributions during the work on the vishaps we thank Alessandra Gilibert and Pavol Hnila, as well as to Pavel Avetisyan,  Karen Bayramyan and Juan Antonio Belmonte for valuable discussions. This study is based on the vishap data collection works supported by the Higher Education and Science Committee of the Ministry of Education, Science, Culture, and Sport (grant no. 21AG-6A080).

**Author contributions**

Both authors wrote the main manuscript text, A.B. prepared Figs. 1,6,7, V.G. prepared Figs. 3, 4. Both authors reviewed the manuscript.

**Competing interests**

The authors declare no competing financial or non-financial interests.

**Additional information**

Correspondence and requests for materials should be addressed to Vahe Gurzadyan.

**Data availability**

All relevant data are contained within the manuscript and archaeological materials are stored at the Institute of Archaeology and Ethnography, National Academy of Sciences of Armenia.